\documentclass[fleqn,usenatbib,usedcolumn]{mnras}
\usepackage[british]{babel}             % British English hyphenation
\usepackage{newtxtext}                  % Good fonts
\usepackage[T1]{fontenc}                % Good font encoding
\usepackage{graphicx}	% Including figure files
\usepackage{amsmath}	% Advanced maths commands
\usepackage{amssymb}	% Extra maths symbols
\usepackage[slantedGreek]{newtxmath}    %   "    "   (slanted Greek)
\usepackage{xcolor}
\long\def\revised#1{{\color{violet}#1}}
\long\def\revised#1{#1}
\DeclareRobustCommand{\VAN}[3]{#2}
\let\VANthebibliography\thebibliography
\def\thebibliography{\DeclareRobustCommand{\VAN}[3]{##3}\VANthebibliography}
\def\chead#1{\multicolumn{1}{c}{#1}}
\def\vpad{\phantom{\Large$|$}}
\hypersetup{pdfauthor={R. E. Spencer et al.},
            pdftitle={Major and Minor Flares on Cygnus X-3 Revisited},
            bookmarksnumbered=true}
\title{Major and Minor Flares on Cygnus X-3 Revisited}

\author[R.~E.~Spencer et al.]{Ralph E.\ Spencer$^{1}$\thanks{E-mail: \texttt{ralph.spencer@manchester.ac.uk}}, Michael Garrett$^{1}$, Justin D.\ Bray$^{1}$
and David A.~Green$^{2}$\\
$^{1}$Jodrell Bank Centre for Astrophysics, Dept.\ of Physics and Astronomy, The University of Manchester, Oxford Rd., Manchester M13 9PL, UK\\
$^{2}$Cavendish Laboratory, 19 J. J. Thomson Ave., Cambridge, CB3 0HE, UK}

\date{Accepted XXX. Received YYY; in original form ZZZ}

\pubyear{2021}

\begin{document}
\label{firstpage}
\pagerange{\pageref{firstpage}--\pageref{lastpage}}

\maketitle

\begin{abstract}
Intense flares at cm-wavelengths reaching levels of tens of Jy have been observed from Cygnus X-3 for many years. This active high mass X-ray binary also has periods of quenching before major outbursts, and has minor flares at levels of a few hundred mJy. In this paper we show that the minor flares have much shorter rise times and durations suggesting more rapid expansion of the synchrotron radiation emitting material than in the strong flares. They also appear closer to the binary, whereas the large flares form a more developed jet. Calculations of physical conditions show that the minor out-bursts have lower minimum power but have larger magnetic fields and energy densities than the major flares. Minor flares can occur while a major flare is in progress, suggesting an indirect coupling between them. The spectral evolution of the minor flares can be explained by either an expanding synchrotron source or a shock model. The possibility that there is a brightening zone as in SS433 is explored.
\end{abstract}



%%%%%%%%%%%%%%%%% BODY OF PAPER %%%%%%%%%%%%%%%%%%

\section{Introduction}\label{sec:intro}

Cygnus X-3 is an X-ray binary system discovered by X-ray detectors on an Aerobee rocket \citep{Giacconi1967}. It is thought to be a $2.4\revised{^{+2.1}_{-1.1}}$~M$_{\sun}$ compact object with a high mass Wolf--Rayet companion \citep{Kerkwijk1992, Kerkwijk1996, Zdziarski2013}. The high activity of the object suggests that the compact object is a black hole though the mass function is uncertain due to high obscuration in the optical band. It has a short binary period (4.8~h) and hence has a close binary orbit ($\sim 3 \times 10^{11}$~cm) with the compact object embedded in an intense stellar wind. The object lies in the Galactic plane at a distance of $7.4 \pm 1.1$~kpc \citep{McCollough2016}. It is noted for its major radio flares at cm wavelengths which rise rapidly and then decay over a few days, reaching levels of 10--20~Jy, \citep{Gregory1972a, Johnston1986, Waltman1995, Fender1997}. The object has been extensively studied with NRAO's Green Bank Interferometer (GBI) and \citet{Waltman1994} found three phases of emission: minor flaring during periods of quiescence at levels of a few 100~mJy, quenching where the flux density at 8~GHz drops below 30~mJy, and major flaring. Quenching occurs for several days before major flares \citep{Waltman1996} and the hard X-ray flux also drops \citep{McCollough1999}, suggesting a close relationship between the accretion disk and the ejection of plasmons in a radio jet.

High resolution MERLIN, VLA and VLBI observations have revealed the presence of relativistic jets with two sided \citep{Marti2001, MillerJones2004} and single sided ejections \citep{Mioduszewski2001, Tudose2010}. Expansion velocities vary from 0.3$c$ \citep{Spencer1986} to 0.81$c$ \citep{Mioduszewski2001} and apparent expansion at 2--3$c$ \citep{Newell1998}. \revised{Apparently s}uperluminal velocities and complex changes in the structure suggest that the jets lie close to the line of sight \citep{Mioduszewski2001, MillerJones2004, Tudose2010}.

The flux density evolution of the flares was originally described by synchrotron self-absorption in an expanding source \citep{Gregory1972b}, where the source is initially optically thick as indicated by a flat or rising radio spectrum with frequency, before expanding and becoming optically thin. The effects of absorption in a stellar wind \citep{Seaquist1977, Fender1997, Miller-Jones2009} in the context of shock-in-jet models have also been considered. % Most of the modelling has been concerned with the major flares; here we examine minor flares in more detail.

Recently \citet{FenderBright2019} have looked into self-absorption and the minimum energy condition in optically thick flares and have derived accurate formulae for the expansion velocity, magnetic field and total energy corresponding to the minimum energy condition for flares in Galactic black hole candidate objects, including Cygnus X-3. In this paper we \revised{extend the application of} these formulae to minor flares and compare the results with those of a selection of optically thick major flares. Observations of the minor flares are described in Section~\ref{sec:flares}. Section~\ref{sec:flares} also lists a selection of major flares. The physical conditions in minor and major flares are compared in Section~\ref{sec:comparison}, followed by discussion in Section~\ref{sec:discussion} and conclusions in Section~\ref{sec:conclusions}.

%%%%%%%%%%%%%%%%%%%%%%%%%%%%%%%%%%%%%%%%%%%%%%%%%%%%%%%%%%%%%%%%%%%%%%
% table of minor flares 

\begin{table*}
\centering
\caption{\revised{Examples} of minor flares. The date, instrument, frequency and UT (in hours) of the peak of the flares is shown together with the peak change in flux density above background, rise and fall times and duration (full width at half maximum) in hours.}\label{tab:minor_flares}
\begin{tabular}{lc......} \hline
	Date        & \chead{Instrument} &\chead{Frequency} & \chead{UTC peak} & \chead{Peak Flux Density} & \chead{$t_{\rm rise}$} & \chead{$t_{\rm fall}$} & \chead{$t_{\rm dur}$} \\
	          &  & \chead{GHz}  & \chead{h} & \chead{Jy} & \chead{h} & \chead{h} & \chead{h} \\\hline
	1983 Oct 7  & MERLIN & 5  & 15.5  & 0.30 & 1    & 1    & 1.5 \\
	1983 Oct 7  & MERLIN & 5  & 22.5  & 0.32 & 2    & 1.3  & 3   \\
	1983 Oct 8  & MERLIN & 5  &  2.5  & 0.15 & 1.5  & 1    & 1.5 \\
	1983 Oct 19 & MERLIN & 5  &  23   & 0.16 & 1.5  & 2    & 2   \\
	1992 Aug 22 & WBI    & 5  & 15.5  & 0.40 & 1    & 3    & 2   \\
	1992 Aug 23 & WBI    & 5  &  2.4  & 0.50 & 1    & 3    & 2   \\
	1995 May 7  & VLBA   & 15 & 11.2  & 0.25 & 0.5  & 1    & 1.5 \\
	1995 May 7  & GBI    & 8.3 & 11.7  & 0.22 & 0.5  & 0.6  & 1   \\
	1995 May 7  & GBI    & 2.25 & 12.6  & 0.05 & 2    & 3    & 2   \\
	1995 May 7  & VLBA   & 15.3  & 15.5  & 0.12 & 0.5  & 0.7  & 0.6 \\
	2020 Feb 14 & AMI    & 15  & 11.39 & 0.11 & 0.3  & 0.23 & 0.5 \\
	2020 Feb 14 & AMI    & 15  & 13.11 & 0.04 & 0.5  & 0.6  & 0.6 \\
	2020 Feb 15 & AMI    & 15  & 11.49 & 0.21 & 0.67 & 0.6  & 1.0 \\
	2020 Feb 15 & AMI    & 15  & 12.61 & 0.10 & 0.3  & 0.5  & 0.5 \\\hline
\end{tabular}
\end{table*}

% Minor flares

\begin{figure*}
\includegraphics[width=\textwidth]{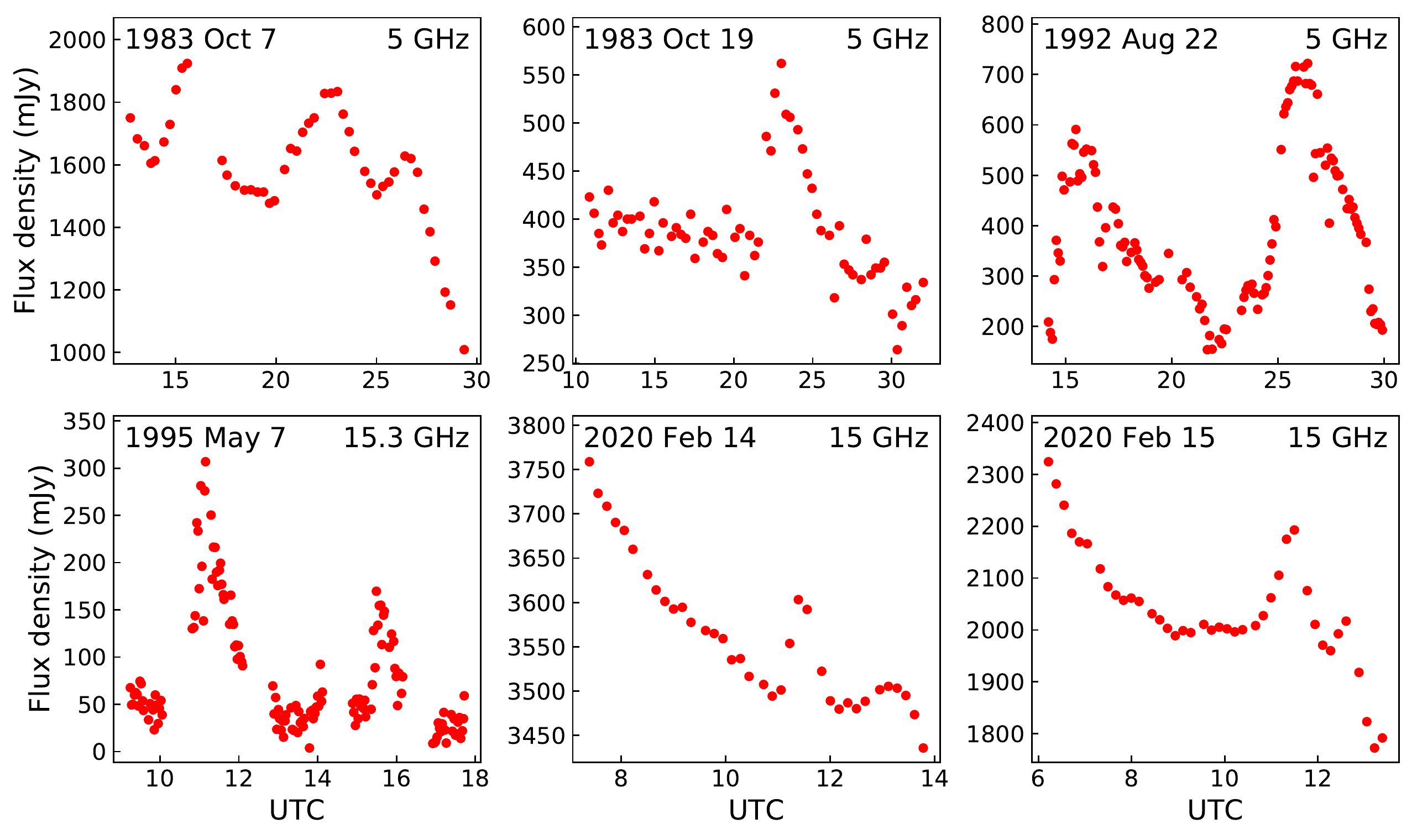}
\caption{Flux density plots for minor flares. The horizontal axis in each case is in hours following 00~h UTC on the date \revised{of the flare, and the frequency is 5~GHz or 15~GHz as shown in the plot.  The flare of 1995 May~7 (lower-left panel) is} also shown in Fig.~\ref{fig:Newfig2}.}\label{fig:MinorFlares}
\end{figure*}

\section{Minor and Major Flares}\label{sec:flares}

\subsection{Minor Flares}\label{sec:minor} % used for referring to this section from elsewhere

Minor flaring (flares with peak flux density between $\sim 0.1$ and 1~Jy) has been has known for some time \citep{Waltman1996} and has been described recently \citep{Egron2021}. Here we look at \revised{examples of} historical data \revised{showing minor flares and possessed by the authors}. The observations are listed in Table~\ref{tab:minor_flares}. The \revised{datasets and corresponding} instruments are as follows\revised{.}
\begin{enumerate}
 \item \revised{The} 1983 data \revised{are from MERLIN} (short baselines only) \citep{Johnston1986}, \revised{triggered by a major flare in September 1983 and observed in a collaborative programme which ran for two months.  The GBI was also used during these observations but did not have short cadences covering the minor flares.}
 \item \revised{The} 1992 data \revised{are from an} unpublished data\revised{set} taken with the Jodrell Bank Mk1--Mk2 wide band interferometer \citep[WBI][]{Padin1987}\revised{, with} observational techniques and calibration \revised{per} \citet{NelsonSpencer1988}.  \revised{This two-day observing run was untriggered.}
 \item The \revised{untriggered} 1995 data are from the VLBA \revised{(short baselines only)} at 15~GHz \citep{Newell1998}, \revised{with corresponding} GBI \revised{data} at 8.3 and 2.25~GHz.
 \item Finally the 2020 data were taken with MRAO's Arcminute Micro\revised{k}elvin Imager (AMI) \revised{triggered by a major flare} \citep{Green2020}.
\end{enumerate}

The flux density time series are shown in Fig.~\ref{fig:MinorFlares}. These flares are quite short in rise time and duration and can be missed in observations with long intervals between measurements.  The data selected were chosen for their short cadences allowing these short lived ($\sim 1$~h) flares to be studied. \revised{Other examples of minor flares have been published \citep[e.g.][]{Ogley2001,Egron2021} but they have similar properties (flux densities, rise times and and duration) to those shown in Table~\ref{tab:minor_flares} and so have parameters which are not significantly different to those calculated here.}  Simultaneous observations at multiple wavelengths are rare, though there are some data available for most flares from the GBI but with longer intervals between data points. The 1983 Oct~7 flare had measurements at 22.78~h and 26.26~h, close to the peaks of the second and third flares at 5~GHz, both with a flat spectral index of $-0.18$ between 8.1~GHz (X-band) and 2.3~GHz (S-band) indicating they were optically thick. \revised{For the flare on 1983 Oct~19, the peak in the 5~GHz emission of $\sim 550$~mJy is roughly simultaneous with a measurement of the 2.3~GHz emission of 65~mJy at 24.06~h,} indicating the flare has a rising spectrum with frequency. The spectral index between X and S band at the onset of the second flare on 1992 Aug~22 is $+0.51$, again showing that this flare is optically thick in the rising stages. \revised{This is important since the \citet{FenderBright2019} analysis assumes that the flare onset is optically thick.}

\begin{figure}
\includegraphics[width=\columnwidth]{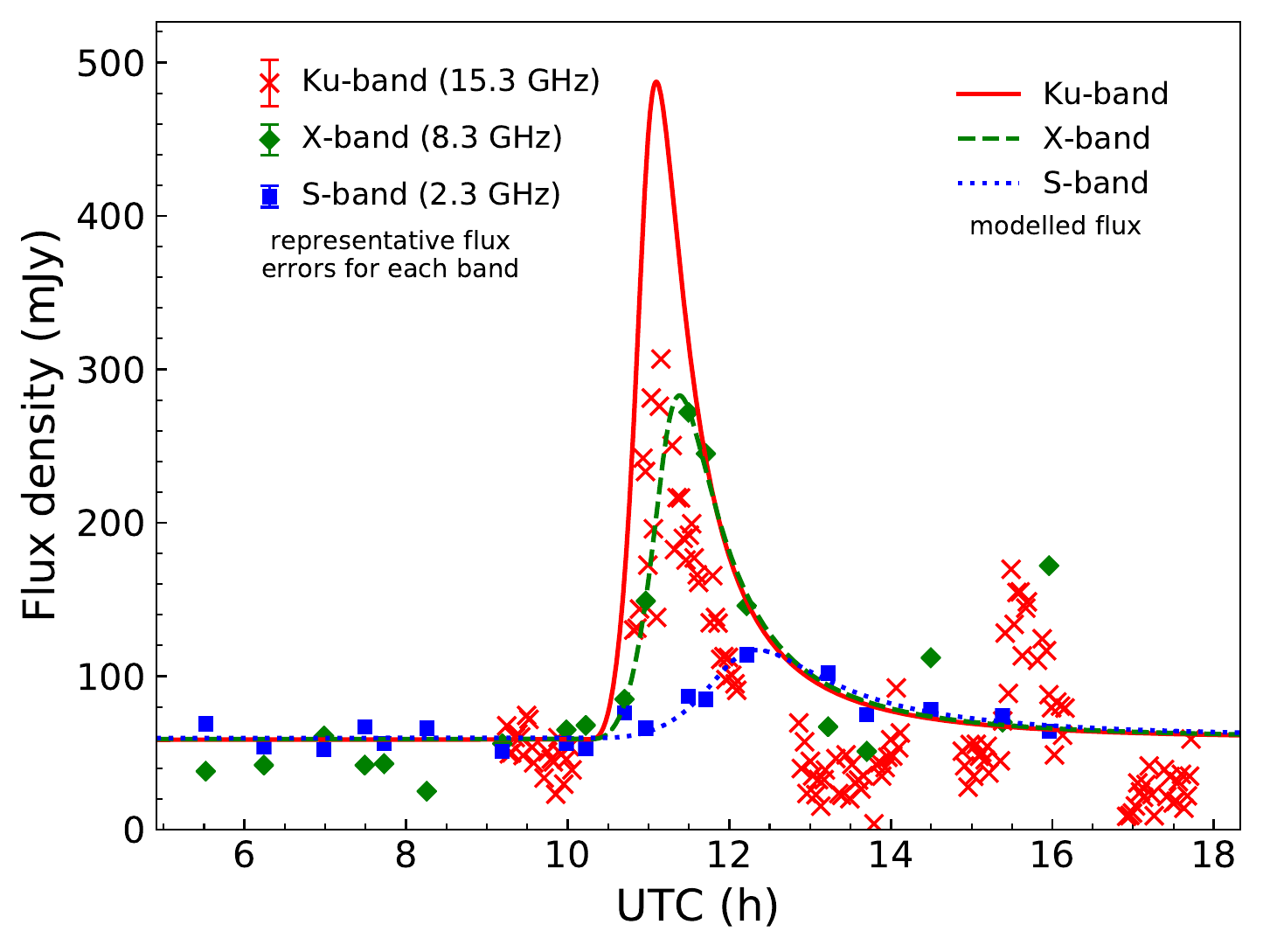}
\caption{Flux densities from the VLBA, and the Green Bank Interferometer on 1995 May~7, at 15.3~GHz, 8.3~GHz and 2.3~GHz.  Also shown are the results of fitting the model of \citet{BallandVlassis1993} to the data (see Section~\ref{sec:discussion}).}\label{fig:Newfig2}
\end{figure}

Luckily the first flare at 15.3~GHz (Ku-band) on 1995 May~7 was well covered by the GBI (Fig.~\ref{fig:Newfig2}) although the second flare at 15.3~GHz is followed by only one data point at X-band with the GBI and is missed entirely at S-band. The data show that the first flare was had a flat or inverted spectrum throughout, only becoming steep spectrum and optically thin after it had decayed at X-band. The peaks at X-band and S-band occur at $0.5 \pm 0.1$ and $1.6 \pm 0.3$~h after the peak at Ku-band.

Simultaneous data at other frequencies for the flares in 2020 Feb are not available. However examination of the flux density across the 12--18 GHz band of AMI allows an estimate of the spectral index. The source was spectrally steep during the decay of a major flare at the time of the minor flares, but the signal to noise ratio was not enough to see if the relatively small changes in flux density due to the minor flaring were flatter.  The rapid rise time and short lived nature of the flares is a good indicator that they are compact with a high brightness temperature and therefore likely to be optically thick at least  in their rising phase.

It is interesting to note that the flares on 1983 Oct~7 and 19 and 2020 Feb~14 and 15 occur during the decay of major flares. \revised{However the observations were triggered by major flaring, so this is not surprising.} The others are not associated with major flares.

\citet{Egron2021} discuss the binary period of 4.8~hours being detectable in the radio. This was first suggested by \citet{Molnar1984} during a period of minor flaring. \citet{Egron2021} also mention the 1995 VLBA measurements of \citet{Newell1998}, noting that the two peaks are $\sim 4.8$~hours apart. However we note that there is evidence of the beginning of an intermediate minor flare at 14.07~UTC, at both 15 and 8.3~GHz (only one data point each), 3 hours after the first peak. The other figures show intervals between peaks which vary from 5 to 11 hours. \revised{Red noise can give rise to apparent quasi periodic behaviour \citep{Vaughan2016} and we do not find strong evidence for the orbital period in our data.}  It is clear that continuous observations with short cadences of 10 minutes or less are needed to follow these short lived flares accurately.

\subsection{Major Flares}\label{sec:major}

A list of major flares taken from the literature \revised{but mostly} using the GBI archive are listed in Table~\ref{tab:major_flares}. \revised{The GBI monitoring programme had 3 observations of Cygnus X-3 each day when operating \citep{Waltman1995}, over the period 1982--1994. There were $\sim$ 20 major flares in this period, several of which were steep spectrum (see the comment by \citet{Fender1997})}. The data for the flare on 1983 Oct~1 were taken with MERLIN \citep{Johnston1986} \revised{since the onset of this flare was missed by the GBI,} those for 2020 Feb~11 with AMI \citep{Green2020}. \revised{The rest are from the GBI: these are the majority of optically thick flares}.  The initial phases of the flares had inverted or flat spectra (as shown by the GBI data \revised{where the X-band flux density is $\geq$ S-band flux density}) and so are optically thick at the start. Examination of the broad-band (12--18~GHz) data from AMI for the 2020 Feb flare shows that the initial phase of the flare had an inverted spectrum.  Only three flares had spectral indices less than 0.0 at the peak of the flare. The flare on 1994 Feb 20 is the first flare in a sequence of 5 discussed by \citet{Fender1997}. They applied equipartition arguments to flare 5 which was optically thin at its peak, whereas flare 1 discussed here had a spectral index of +1.69 at its peak and so is optically thick.

The flux densities of the major flares, by definition, are much greater than for the minor flares but also their rise times and duration are much longer -- days rather than hours. This has implications for the energy required to generate the flares. 

% table of major flares

\begin{table}
\centering
\caption{\revised{Examples} of major flares. The date,frequency and change in flux density at the peak of the flares is shown together with rise and fall times and duration (full width at half maximum) in days.}\label{tab:major_flares}
\begin{tabular}{l.....}\hline
Date & \chead{Frequency} & \chead{Peak Flux} & \chead{$t_{\rm rise}$} & \chead{$t_{\rm fall}$} & \chead{$t_{\rm dur}$} \\
     &                   & \chead{Density} & & & \\
	 & \chead{GHz} & \chead{Jy} & \chead{d} & \chead{d} & \chead{d} \\\hline
1983 Oct 1  &  5   & 14.5 & 0.25 & 4    & 3    \\
1983 Oct 11 &  8.1 &  9.0 & 1.0  & 3    & 3    \\
1985 Oct 10 &  8.1 & 13.0 & 1.25 & 3    & 3    \\
1985 Dec 30 &  8.1 &  8.5 & 1.5  & 1.8  & 4.5  \\
1989 Jun 2  &  8.1 & 16.8 & 1.25 & 4.2  & 2    \\
1989 Jul 21 &  8.1 & 17.5 & 1.0  & 3    & 2    \\
1990 Aug 15 &  8.3 &  7.4 & 1.0  & 2.5  & 2    \\
1990 Oct 5  &  8.3 &  9.0 & 0.75 & 2.9  & 1.5  \\
1991 Jan 21 &  8.3 & 14.0 & 1.0  & 3    & 1.25 \\
1991 Jul 26 &  8.3 & 18.2 & 0.5  & 0.75 & 0.7  \\
1994 Feb 20 &  8.3 &  5.0 & 1.0  & 2    & 1    \\
2020 Feb 7 & 15   & 9.9 & 0.2  & 0.25  & 0.25 \\\hline
\end{tabular}
\end{table}

\section{Comparison of major and Minor Flare Minimum Energy conditions}\label{sec:comparison}

Calculations assuming equipartition between the energy in the magnetic field and in relativistic electrons have often been used as an indicator of the energy and magnetic field \citep{Miley1980, Spencer1984, Longair2011} in radio sources. However there is no guarantee that flares in objects like Cygnus X-3 are in equipartition -- in fact the rapid variability of an expanding source implies otherwise. A more useful condition is minimum energy since that sets a lower limit to the energy in the source, though in fact the values are close to those found assuming equipartition \citep{Fender2006, Longair2011}.

Magnetic fields can also be found for a source showing synchrotron self-absorption, for example \citet{MillerJones2004}, however the results are very strongly dependant on the angular diameter $\theta$ of the object ($B\propto \theta^4$). Assuming that flares are initially optically thick in the injection phase they will reach peak flux density as the source expands and the optical depth to synchrotron self-absorption becomes close to $\sim 1$. Adiabatic expansion losses result in a decrease in flux density as the source becomes optically thin. \citet{FenderBright2019} have produced practical formulae which combine self-absorption and the minimum energy condition and which have relatively weak dependencies on the observed parameters.  They find that the total energy is a function of an effective expansion rate  $\beta=f^{1/3}v/c$ where $f$ is the filling factor. The maximum size occurs for $f=1$ and $v=c$ whereas a minimum size is set by the Compton limit for a brightness temperature of $\sim 10^{12}$~K. A uniform spherical source is assumed in their calculations. Solving for minimum energy gives an estimate for $\beta$. Note that the contribution from baryons in the source has been neglected in these calculations so a true lower limit to the  energy can be found.

Their approximate formulae for the single frequency case assuming optically thick flares (their equations 28--31) are used, and calculated assuming a distance of 7.4~kpc. The equations give values for the brightness temperature $T_{\rm B}$, the effective expansion velocity $\beta$, the minimum total energy $E$ in the flare and the magnetic field $B$. Their formulae, omitting constants and the distance term, are given in equations (1) to (4). 
\begin{gather}
  T_{\rm B} \propto F^{1/17} \nu^{-1/17} \\
  \beta \propto F^{8/17} \nu^{-33/34} t_{\rm dur}^{-1} \\
  E \propto F^{20/17} \nu^{-23/34} \\
  B \propto F^{-2/17} \nu^{19/17} 
\end{gather}
The weak dependence of the brightness temperature $T_{\rm B}$ on the maximum increase in flux density $F$ above background in the flare and the observing frequency $\nu$ can be seen. Note that $F$ is used here following \citeauthor{FenderBright2019}, rather than the more conventional total flux density $S$ since the change in flux density above background is used. In addition the energy density
\begin{equation}
    U=\frac{E}{V}
\end{equation}
and the power $P$ required to generate the flare, assuming injection occurs in the rising phase,
\begin{equation}
  P=\frac{E}{t_{\rm rise}}     
\end{equation}
have been calculated for each of the flares listed in Table~\ref{tab:minor_flares} and Table~\ref{tab:major_flares}. We assume spherical geometry and volume
\begin{equation}
  V=\frac{4\uppi}{3} (\beta c t_{\rm dur })^3
\end{equation}
since the plasmon will expand through the duration of the flare, i.e.\ $\theta=\beta c t_{\rm dur}$. Only the 15.3~GHz data were included in the calculation for the 1995 May minor flares. Table~\ref{tab:calculations} shows the average values for the parameters calculated from equations (1) to (7).

\begin{table*}
\centering
\caption{Average minimum energy parameters calculated for minor and major flares. The standard deviation of the mean of the values is also shown.}\label{tab:calculations}
\begin{tabular}{lccc..c}\hline
     & $T_B$ & $\beta$ & $E$ & \chead{$B$} & \chead{$U$} & $P$\\
     & K & & erg & \chead{gauss} & \chead{erg~cm$^{-3}$} & erg~s$^{-1}$ \\\hline
Minor Flares\vpad & $5.3 \times 10^{10}$  & 0.13  & $2.8 \times 10^{39}$ & 1.2  & 0.22  & $8.5 \times 10^{35}$ \\
$\sigma_m$   & $1.0 \times 10^{9}$   & 0.01  & $7.6 \times 10^{38}$ & 0.2  & 0.06  & $1.9 \times 10^{35}$ \\
Major Flares & $6.75 \times 10^{10}$ & 0.033 & $2.5 \times 10^{41}$ & 0.58 & 0.044 & $5,0 \times 10^{36}$ \\
$\sigma_m$   & $6.4 \times 10^{8}$   & 0.007 & $3.8 \times 10^{40}$ & 0.05 & 0.01 & $1.6 \times 10^{36}$ \\\hline
\end{tabular}
\end{table*}

It can be seen that the minor flares have lower brightness temperatures, higher expansion velocities, lower total energy, higher magnetic fields and energy densities and lower powers under the assumption of minimum energy in comparison with the major flares. These parameters are compatible with the emitting region being more rapidly varying and hence more compact than those in the major flares. There is a narrow spread in brightness temperature as expected from the weak dependence on flux density, however major flares are $\sim 100$~times stronger in flux density and hence are significantly brighter, There is a wide spread in the minimum power for both types of flare, due to the wide variation in flux density and rise time, but on average the major flares are more powerful, even though their rise times are much longer.

\subsection{Doppler Effects}

Bulk motion of the radiating region could result in significant Doppler factors:
\begin{equation}
    \delta=\frac{1}{\Gamma (1 \mp \beta_{\rm b} \cos i)}
\end{equation}
where $\Gamma$ is the Lorentz factor, $\beta_{\rm b}$ the bulk velocity in units of $c$, to be distinguished from the effective expansion velocity above, and $i$ the angle to the line of sight.

\citet{FenderBright2019} point out that Doppler boosting with $\delta >$1 will result in the intrinsic values being smaller than calculated.  A variety of Doppler factors in major flares have been estimated for Cygnus X-3 and the structural changes observed suggest the jet lies close to the line of sight \citep[see e.g.][]{Tudose2010}. VLBI imaging of this source is notoriously difficult due to its rapid variability.  One of the most thorough evaluations has been by \citet{MillerJones2004} where $\beta_{\rm b} = 0.56$ and $i = 25\fdg7$ giving $\delta=1.83$ for a distance of 10~kpc. The source size was $\sim 40$~mas. Scaling to a distance of 7.4~kpc gives $\delta=1.70$. Since the minimum energy is $\propto\delta^{97/34}$, the power is $\propto \delta^{131/34}$, giving factors of 4.5 and 7.7 respectively if the component is approaching. The rest frame powers are therefore a factor of 7.7 smaller than those given in Table~\ref{tab:calculations} and well below the Eddington luminosity for this source.

We do not know if the minor flares, which appear on sub-mas size scales, have significant Doppler factors. The observations by \citet{Newell1998} suggest super-luminal expansion at $\sim 2c$. This requires $\beta_{\rm b} >0.9$, $i < 25 \degr$ and a Doppler factor $\delta >2.3$. We would also perhaps expect the source to be moving more quickly at early stages if it is eventually slowed down by interaction with the surrounding medium, though there is no significant evidence for this. The major flares would need to be significantly Doppler boosted by more than observations suggest for their rest frame energies and powers to become comparable with those of minor flares.

%%%%%%%%%%%%%%%%%%%%%%%%%%%%%%%%%%%%%%%%%%%%%%%%%%%%%%%%%%%%%%

\section{Discussion}\label{sec:discussion}

A minor flare observed with the Ryle telescope on 2003 Jan~5 at 15~GHz was described by \citet{Fender2006}. The 200~mJy flare rose in $\sim 1$~h and with total energy of $5 \times 10^{40}$~erg and power of $\sim 10^{37}$~erg~s$^{-1}$ assuming equipartition. This is more powerful than the values found here, but consistent with the minimum energy assumption. The magnetic field was 0.5~gauss, similar to values here. Recently \citet{Broderick2021} observed a self-absorbed major flare with LOFAR, with a much greater power of $\sim 10^{38}$~erg~s$^{-1}$, due to the lower radio frequency of the observations. This is still less than the Eddington luminosity of $3.0 \times 10^{38}$~erg~s$^{-1}$ for a 2.4~M$_{\sun}$ compact object. However the major flares do seem to need a greater power, and this is consistent with the formation of an extended jet in these events, where the extra power is needed to propel the emitting plasma further from the compact object. The minor flares have lower power but expand more rapidly and cannot expand out very far before decaying. They are therefore likely to be closer to the compact object than much of the emitting region in the major flares.

Comparison with X-rays \citep{Szostek2008} shows that the radio vs X-ray luminosity plot has a reversed `h' shape. As the X-ray flux increases the radio moves from quiescence to minor flaring, and back to quiescence if the X-ray flux decreases, but can also move to the quenched state if the X-ray flux increases preceding a major radio flare. After the flare the source reverts to a lower X-ray state and minor flaring can occur again. Our observations show that minor flaring can occur while the major flare is decaying, but it is likely that the radio jet emission by this stage has moved away from the core, and the compact object/disk system has reverted to the lower X-ray state. Detailed high cadence X-ray observations during minor and major flares are needed to confirm this. 

There are three basic models to describe the spectral evolution of Cygnus X-3 flares (see e.g.\ \citealt{Fender1997}): the expanding plasmon model, as originally described by \citet{vanderLaan1966} and modified by \citet{HjellmingJohnston1988} and \citet{BallandVlassis1993}, absorption in a stellar wind \citep{Seaquist1977,Fender1997} and shock-in-jet models \citep{Marti2001, Lindfors2007, Miller-Jones2009}. The 1995 May~7 data have good frequency coverage so can be compared with models.

\subsection{Expanding Plasmon Model}

The expanding plasmon model relies on adiabatic expansion to cause the decrease in brightness and hence transition to becoming optically thin. Fits to a major flare by \citet{Fender1997} suggest that the decay in flux density also occurs via energy losses. Loss of energy by synchrotron radiation cannot explain the decay time-scale, and inverse Compton (IC) losses in the radiation field of the Wolf--Rayet star were suggested. With that in mind, we have fitted a model based on \citet{BallandVlassis1993} (which has adiabatic losses only) to the 1995 May~7 data.

Following \citeauthor{BallandVlassis1993}, we take the flux at time $t$ and frequency $\nu$ to be
\begin{equation}
 S = S_{\rm q} \left(\frac{\nu}{\nu_0}\right)^\alpha +
  \begin{cases}
   S_0 \left(\frac{t - t_{\rm i}}{t_0}\right)^3 \left(\frac{\nu}{\nu_0}\right)^{5/2} (1 - {\rm e}^{-\tau}) \, \xi_3(\tau) & t \geq t_{\rm i} \\
   0 & t < t_{\rm i}
  \end{cases}
\end{equation}
where
\begin{equation}
 \xi_3(\tau) = \frac{1 - 2 \left( 1 - (1 + \tau) {\rm e}^{-\tau} \right) \tau^{-2}}{1 - e^{-\tau}}
\end{equation}
and
\begin{equation}
 \tau = \tau_0 \left(\frac{t - t_{\rm i}}{t_0}\right)^{-(2\gamma+3)} \left(\frac{\nu}{\nu_0}\right)^{-(\gamma+4)/2} ,
\end{equation}
in which we have modified the original model by inserting an initial time offset $t_{\rm i}$ and a background quiescent flux with scale factor $S_{\rm q}$.  Taking $\nu_0 = 15.3$~GHz, we find best-fit values for the remaining parameters with a global optimiser from \citet{Storn1997}, fitting modelled fluxes $S_{\rm mod}$ to observed fluxes $S_{\rm obs}$ by minimising the metric $\sum_{\nu,t} \left( S_{\rm mod}(\nu,t) - S_{\rm obs}(\nu,t) \right)^2 / \sigma_\nu^2$, with results listed in Table~\ref{tab:model}.  Most of the parameters are well-constrained, but there is considerable degeneracy in the values of $\tau_0$ and $S_0$; however, the product $\tau_0^3 \, S_0^{2\gamma+3}$ is well constrained, and the degeneracy has little affect on the resulting physical flux $S$.  The resulting modelled flux is shown alongside the observational data in Fig.~\ref{fig:Newfig2}.

\begin{table}
 \centering
 \caption{Parameters of the modified \citet{BallandVlassis1993} model (see text) fitted to the VLBA and GBI data of 7 May 1995.}
 \label{tab:model}
 \begin{tabular}{cc}\hline
  \chead{Parameter} & \chead{Value} \\ \hline
  $t_{\rm i}$ & 10.25~h \\
  $t_0$ & 0.85~h \\
  $\tau_0$ & 1.493 \\
  $\gamma$ & 1.224 \\
  $S_0$ & 708.26 mJy \\
  $S_{\rm q}$ & 58.60 mJy \\
  $\alpha$ & $-0.010$ \\
 \hline
\end{tabular}
\end{table}

The emission does not become optically thin until after the peak at each frequency but then decays rapidly. Energy losses other than adiabatic (synchroton and IC) are not needed to explain the rapid decay. However the model does not fit the 15.3~GHz data well, with the observed flux density being much less than predicted by the model. This implies that extra energy losses are needed to explain a break in the spectrum. If the break occurs at $\sim 10$~GHz then the electron lifetime for synchrotron energy losses for a magnetic field of 1.2~gauss is $\sim 3$ months, so this can be ruled out. A high energy density is required to give the short ($<$1~h) timescale required for IC losses. Such losses could occur in the \emph{uv} radiation field of the Wolf--Rayet companion, as is the case for the major flares. For a star of luminosity $10^{39}$~erg~s$^{-1}$ the energy density is high enough for this to happen for a radius from the star of $\sim 10^{12}$~cm for a break in the radio spectrum at 10~GHz.  This is relatively close to the star and comparable with the binary orbit so the losses must occur at an early stage of the initial injection of relativistic plasma. The \emph{uv} photons would be up-scattered into $\gamma$ rays \citep{Piano2012}. Recent results from Fermi-LAT \citep{Trushkin2017} and AGILE \citep{Piano2021a,Piano2021b} $\gamma$ ray satellites show $\gamma$ emission during minor flaring in the radio when just before the quenched state and also at the onset of major flaring at the end of the quenched state. \citet{Koljonen2018} suggest this might arise from shocks in the jet, but our results indicate that the $\gamma$ ray emission could also be from IC interactions of the high energy electrons responsible for the radio synchrotron emission at the injection phase of the plasmon or jet evolution.

If expanding at 0.13$c$ (Table~\ref{tab:calculations}) and with an average duration of 1.4~h then the plasmon must be compact ($\sim 10^{13}$~cm). Given an average flux density for the minor flares of 0.22~Jy and brightness temperature of 5.3$\times$10$^{10}$~K the plasmon is $\sim 0.16$~mas in angular size or $\sim 2 \times 10^{13}$~cm in diameter. The energy density in the Wolf--Rayet radiation field would be less at this distance and so IC losses less important as the plasmon expands.

\subsection{Free--free Absorption}

An alternative model is to assume the spectral evolution is due to free--free absorption in a strong stellar wind. The longer time-scale at lower frequencies in the 1995 May~7 data is a good indicator that opacity effects are important. The electron density in a simple spherically symmetric wind, assuming mass conservation and a uniform velocity $v$ is
\begin{equation}
  N= \frac{\dot{M}} {4\pi \mu m_{\rm p} v r^2}
\end{equation}
where $\dot{M}$ is the mass loss rate, $\mu$ is the mean molecular weight, $m_{\rm p}$ the mass of the proton and $r$ the radius from the Wolf--Rayet companion. The optical depth in the wind from radius $R$ is
\begin{equation}
   \tau = \int_{R}^{\infty} \kappa_\nu N^2 \, {\rm d}r 
\end{equation}
where $\kappa_\nu$ is the absorption coefficient at frequency $\nu$ and so $\tau$ is $\propto \dot{M}^2 R^{-3}$.
 
The radius at which the optical depth is unity can then be found, using the formulae for absorption coefficient and Gaunt factor in \citet{Spitzer1978}. For a hot intense wind at $10^{5}$~K, $\mu = 0.62$ (solar value), $v= 1000$~km~s$^{-1}$ and a mass loss rate of $10^{-5}$~M$_{\sun}$~yr$^{-1}$ the unity optical depth radii are $1.3, 1.9, 4.7 \times 10^{13}$ cm at 15.3, 8.3 and 2.25~GHz respectively. These are all much larger than the orbital diameter. If a moving source of emission becomes visible when the optical depth drops to approximately unity, then the region must be moving at $\sim 0.1$--$0.3c$, comparable with the expansion velocity for minor flares in Table~\ref{tab:calculations}. However the radio spectrum of the emitting region must change as it moves out, since there are rapid decreases in flux density and a clear transition from optically thick with an inverted spectrum to eventually becoming optically thin. The higher frequency emission decreases rapidly in the observations, whereas we would expect the standard synchrotron spectrum when the source becomes optically thin. So motion through a wind does not fully explain the spectral evolution of the flare, \revised{though we expect the lower high-frequency emission caused by inverse Compton losses in the early phases of the flares to also be visible as the plasmon moves outward.}

The energy density $U$ is high (0.22~erg~cm$^{-3}$) for the minor flares and therefore the plasmons have higher pressure than that in major flares. The internal pressure is equal to the gas pressure in the hot wind at a distance of $1.8 \times 10^{12}$~cm, and greater than that in the wind at distances beyond that. Rapid expansion is therefore possible at greater distances from the core and could even reach the value for the sound velocity in a relativistic gas of $c/\sqrt3$. 

\subsection{Shock-in-Jet Models}

Shock-in-jet models have been fitted to Cygnus X-3 flares by a number of authors. \citet{Miller-Jones2009} showed that a conical jet model can be fitted to minor flares, though their data had multiple overlapping events. The model also assumed IC and synchrotron losses are important in the rise phase of the flare, as shown by \citet{Lindfors2007} for a major flare on Cygnus X-3. However infra-red data are needed to constrain the model during the rising phase and expanding plasmon models cannot be excluded. The decay rate for spherical expansion is $\propto t^{-2s}$; a slower rate is expected for 2-D expansion in a conical jet and for the shock-in-jet model, though turbulence can also slow the decay rate \citep{Miller-Jones2009}. The fall times of the minor flares in the paper are comparable with the rise times and are consistent with the plasmon model. It is also interesting that \citet{Lindfors2007} also found that a trumpet shaped jet was needed to account for the early phase of flare evolution. The short fall times of the minor flares suggest that a rapidly widening jet is needed.

\citet{Koljonen2018} suggest that any channel formed by previous jet ejections is filled in by the wind from the Wolf--Rayet star during the quenched state. The next outburst results in a strong radio jet with shocks formed by interaction with the wind material. Minor flaring could correspond to jets which fade rapidly because of low density and weak shocks in the channel which has yet to be filled in by the wind.

Assuming that the jet velocities found for the major flares (0.3--0.8$c$) are the same as those for the minor flares, and given the expansion velocities in  Table~\ref{tab:calculations}, then the opening angles of the jets can range from $9^\circ$ to $24^\circ$ for minor flares and $1^\circ$ to $5^\circ$ for major flares. The longer duration (by factors of $\sim 30$) of the major flares also implies they become longer, so minor flares are short and fat, major flares long and thin.

Finally, \revised{several} of the flares in Fig.~\ref{fig:MinorFlares} are double, which leads to the possibility that the emission is seen from the approaching and receding parts of the jet when the jet brightens at a given distance from the central engine. Such a brightening zone was seen in SS433 \citep{Vermeulen1987} at a distance of $\sim 300$ au from the core. \revised{The approaching component appears first and should be brighter than the receding one, unless $\beta_{\rm b}$ is small. Most of the  double flares in Fig.~\ref{fig:MinorFlares} have a brighter first flare}. The time delay, if the brightening occurs at a distance $d$ on each side of the core, is $(2d/c)\cos i$ where $i$ is the angle to the line of sight. The time difference between the peaks varies from $\sim 1$ to 10~h, suggesting either changes in $i$ or in $d$ from epoch to epoch. The images in the VLBI data of \citet{Newell1998} show a shift of $1.0 \pm 0.2$~mas between the peaks of the first and second flares in the 1995 data, though these observations may be limited by strong interstellar scattering. The 4.3~h delay between the peaks implies an angle to the line of site of $13^\circ \pm 2^\circ$ and $d = 16 \pm 1$~au. If the angle to the line of sight is the same for each epoch then the emission peaks at a distance which varies from 3.6 to 36~au from the core. This is much smaller than for SS433, however the binary orbit is also much smaller. \citet{Vermeulen1987} suggested that either a re-collimation shock, interaction of successive ejections with a bow shock, or a working surface in a dense wind occurred at the brightening zone, after which the components expanded adiabatically and  similar effects could be occurring here.

%%%%%%%%%%%%%%%%%%%%%%%%%%%%%%%%%%%%%%%%%%%%%%%%%%%%%%%%%%%%%%%%%%%%

\section{Conclusions}\label{sec:conclusions}

Historic data on minor and major flares have been used to derive physical parameters assuming self-absorption and minimum energy. The much more rapid but weaker minor flares have higher expansion velocities, magnetic fields and energy densities but lower brightness temperatures, total energies and powers than major flares. A rapidly expanding self-absorbed synchrotron model can explain the minor flares, and the emitting plasmons are likely to be formed relatively close to the compact object, whereas the bulk of the emission in the major flares occurs further out. If produced by a shock in a continuous jet then the jet has to be expanding rapidly, and the high energy density implies that the plasmons are not confined by a hot wind from the Wolf--Rayet companion. The plasmons have expanded to $\sim 1$~au after $\sim 1$~h and are at a few times that in distance from the compact object and disk when they become visible at cm-wavelengths, whereas the jets in major flares are $\sim 30$~times longer.

Further investigation of the radio emission requires simultaneous observations at infra-red and millimetre wavelengths to help distinguish plasmon or shock-in-jet models, and X-ray observations will help understand how the flares are initiated. If inverse Compton losses in the radiation field of the Wolf--Rayet star are important for minor flares then the flares may also be observed in $\gamma$ rays. The short time-scale of the flares require short cadences of $\sim$ minutes and observation periods of several hours to fully capture the flares. Finally snap-shot VLBI observations at frequencies >15~GHz are needed to overcome the strong inter-stellar scattering to Cygnus X-3 and the rapid variability to image a potential brightening zone.

\section*{Acknowledgements}

The Green Bank Interferometer (GBI) is a facility of the National Science Foundation operated by the National Radio Astronomy Observatory. From 1978--1996, it was operated in support of USNO and NRL geodetic and astronomy programmes; after 1996, in support of NASA High Energy Astrophysics programmes. We thank the staff of the Mullard Radio Astronomy Observatory, University of Cambridge, for their support in the maintenance, and operation of AMI. We also thank Prof.\ Richard Schilizzi and the referee for their valuable comments on the manuscript.

%%%%%%%%%%%%%%%%%%%%%%%%%%%%%%%%%%%%%%%%%%%%%%%%%%
\section*{Data Availability}

The Green Bank data are available on the NRAO GBI archive under 2030$+$407 (\url{https://www.gb.nrao.edu/fgdocs/gbi/arcgbi/}, last accessed 2021 March~22). The data from other instruments in the form of spreadsheets are available on request from the corresponding author.


\bibliographystyle{mnras}
\bibliography{XRB} 

% Don 't change these lines

\bsp % typesetting comment
\label{lastpage}
\end{document}